\documentclass[prb,aps,twocolumn,showpacs]{revtex4}
\usepackage{color}
\usepackage{amssymb}
\usepackage{amsmath}
\usepackage{amsfonts}
\usepackage{tabularx}
\usepackage{pifont}
\usepackage{makeidx}
\usepackage[dvips]{graphicx}
\usepackage{graphics}
\begin{document}
\bibliographystyle{apsrev}

\newcommand {\abs}[1]{| #1 |}
\newcommand {\bra}[1]{\langle \: #1 \: |}
\newcommand {\ket}[1]{| \: #1 \: \rangle}
\newcommand {\unity}{{\textrm{1}\hspace*{-0.55ex}\textrm{l}}}
\newcommand {\dotp}[3]{\langle \: #1 | \: #2 \: | \: #3 \: \rangle}
\newcommand {\expect}[1]{\langle \: #1 \: \rangle}
\newcommand {\ybco}[1]{YBa$_2$Cu$_3$O$_{#1}$}
\newcommand {\lacuo}{La$_2$CuO$_4$}
\newcommand {\cuo}{CuO$_2$}
\newcommand {\K}[3]{^{#1}\!K_{#2}^{#3}}
\newcommand {\Kbar}[3]{^{#1}\overline{K}_{#2}^{#3}}
\newcommand {\R}[3]{^{#1}\!R_{#2}^{#3}}
\newcommand {\Rbar}[3]{^{#1}\overline{R}_{#2}^{#3}}
\newcommand {\s}[3]{^{#1}\sigma_{#2}^{#3}}
\newcommand {\ybc}{YBa$_2$Cu$_4$O$_8$}
\newcommand {\Cu}{$^{63}$Cu}
\newcommand {\Tone}{$^{63}$T$_1$}
\newcommand {\Tonec}{$^{63}$T$_{1c}^{-1}$}
\newcommand {\nd}{Nd$_2$CuO$_4$}
\newcommand {\ndce}{Nd$_{2-x}$Ce$_x$CuO$_4$}
\newcommand {\lacuosr}{La$_{2-x}$Sr$_x$CuO$_4$}
\newcommand {\infl}{SrCuO$_2$}
\newcommand {\infd}{Sr$_{1-x}$La$_x$CuO$_2$}
\newcommand {\srcl}{Sr$_2$CuO$_2$Cl$_2$}
\newcommand {\srf}{Sr$_2$CuO$_2$F$_2$}

\newcommand {\cuf}{Cu$_{5}$}
\newcommand {\cun}{Cu$_{9}$}
\newcommand {\cut}{Cu$_{13}$}

\newcommand {\dx}{3d$_{x^2-y^2}$}
\newcommand {\dz}{3d$_{3z^2-r^2}$}

\title{First principles study of local electronic and magnetic properties in pure and electron-doped Nd$_2$CuO$_4$}

\author{C. Bersier, S. Renold, E. P. Stoll, and P. F. Meier}

\affiliation{Physics Institute, University of Zurich, CH-8057 Zurich, Switzerland}
\date{\today}
\begin{abstract}
The local electronic structure of \nd\ is determined from ab-initio cluster calculations in the framework of density functional theory. Spin-polarized calculations with different multiplicities enable a detailed study of the charge and spin density distributions, using clusters that comprise up to 13 copper atoms in the CuO$_2$ plane. Electron doping is simulated by two different approaches and the resulting changes in the local charge distribution are studied in detail and compared to the corresponding changes in hole doped \lacuo. The electric field gradient (EFG) at the copper nucleus is investigated in detail and good agreement is found with experimental values. In particular the drastic reduction of the main component of the EFG in the electron-doped material with respect to La$_2$CuO$_4$ is explained by a reduction of the occupancy of the $3d_{3z^2-r^2}$ atomic orbital. Furthermore, the chemical shieldings at the copper nucleus are determined and are compared to results obtained from NMR measurements. The magnetic hyperfine coupling constants are determined from the spin density distribution.
\end{abstract}
 
\maketitle


 \section{Introduction}
Soon after high-T$_c$ superconductivity was discovered~\cite{bib:Bednorz,bib:Wu1987} in hole-doped La$_{2-x}$(Ba,Sr)$_x$CuO$_4$ and YBa$_2$Cu$_3$O$_{7-\delta}$ materials, it was found~\cite{bib:Tokura} that electron-doped Nd$_{2-x}$Ce$_x$CuO$_4$ also superconducts. All these substances contain CuO$_2$ planes separated by block layers containing rare-earth ions, and, moreover, the parent compounds are insulating and exhibit quasi two-dimensional antiferromagnetism. The generic phase diagrams of electron- and hole-doped cuprates, however, are somewhat different and are not totally symmetric with respect to hole or electron doping. This is not unexpected since there are quite significant structural differences. In La$_2$CuO$_4$, the copper is six-fold coordinated with oxygens whereas in Nd$_2$CuO$_4$, the apical oxygens are missing. To gain insight into the mechanism(s) of high-T$_c$ superconductivity, investigations on the origin of similarities and differences between hole- and electron-doped materials are very important.

In a simplistic ionic picture the valencies of the Cu and the O are the same in the insulating state of the parent compounds La$_2$CuO$_4$ and Nd$_2$CuO$_4$. The former substance becomes metallic upon hole doping and it is assumed that the doped hole goes to the planar oxygen sites. On the other hand, in \nd, the metallic state is achieved by electron doping and the doped electrons are assumed to reside on the copper sites.

Electronic structure calculations~\cite{bib:massida1989} in the framework of the full-potential linearized augmented-plane-wave method within the local density approximation have shown very similar band structures for Nd$_2$CuO$_4$ and La$_2$CuO$_4$. Details of the influence of electron or hole doping in these materials, however, cannot be obtained from band structure calculations except in the approximation of rigid bands. In contrast, cluster calculations are ideally suited for investigations of how the local electronic structure is influenced by doping. Using two different methods of simulating doping, we have previously shown~\cite{bib:stollI2003} that simulations of hole doping in La$_2$CuO$_4$ provide insight into the changes in the population of various orbitals on the copper and oxygen atoms. Both methods yielded on the whole the same results and the changes calculated for the electric field gradient (EFG) at the copper agreed with experiments. Parts of this work deal with an adaptation of these methods to the case of the electron-doped cuprate \nd.

A large body of data obtained by nuclear quadrupole resonance (NQR) and nuclear magnetic resonance (NMR) gives detailed information on the changes of the EFG values in all high T$_c$ superconductors. Several NMR experiments were performed in the electron-doped \nd\ compound~\cite{bib:kumagai1989,bib:abe1989,bib:zheng1989,bib:yoshinari1990,bib:kambe1991,bib:kumagai1991}, the Pr$_{1-x}$LaCe$_x$CuO$_{4-\delta}$ material~\cite{bib:zheng2003}, the infinite layer compound~\cite{bib:imai1995,bib:williams2002el,bib:verkhovskii2003} \infd, and its parent compound~\cite{bib:mikhalev2004} \infl. The first surprising feature of these results is a relatively low $^{63}$Cu NQR frequency ($\nu_Q$) measured in the undoped parent compound compared to the hole doped compounds. In the undoped parent compounds the quadrupolar frequency has been measured to be 14~MHz for the the \nd\ compound and 7.4 MHz in the infinite layer. This is less by more than a factor of two than in \lacuo\ (33.0 MHz (Ref.~\onlinecite{bib:imai1993})) and \ybco{7}\ (31.5 MHz (Ref.~\onlinecite{bib:pennington1989})) compounds. The second striking feature of all these experiments is the extreme sensitivity of $\nu_Q$ on the doping level. $\nu_Q$ is $\le$ 2 MHz for Nd$_{1.85}$Ce$_{0.15}$CuO$_4$ (Ref.~\onlinecite{bib:kambe1991}), $\le$ 0.5 MHz in Pr$_{0.91}$LaCe$_{0.09}$CuO$_{4-\delta}$ (Ref.~\onlinecite{bib:zheng2003}) and finally  $\le$ 3 MHz in Sr$_{0.9}$La$_{0.1}$CuO$_2$ (Ref.~\onlinecite{bib:williams2002el}). This strong doping dependence of the field gradient in the electron-doped compounds has been explained by a simplistic ionic model. It was shown to be due to fully occupied 3d$_{x^2-y^2}$ orbitals which give no contribution to the field gradient if all 3d orbitals are occupied.

In the present work we report on cluster calculations of the local electronic structure in Nd$_2$CuO$_4$ and its changes upon electron doping. The results are compared to those for hole-doped La$_2$CuO$_4$. The contributions to the EFG are studied in detail and it is explained why the EFG values in Nd$_2$CuO$_4$ are much smaller than in the hole-doped substances.

In Sec.~\ref{sec:clustermodel} we present the details of the technique of the cluster calculations. In Sec.~\ref{secIII} the results for the electronic structure including the charge distribution, the EFG and their comparison to NMR experiments are given. Especially, in Sec.~\ref{sec:orbital}, results for the chemical shieldings at the copper sites in undoped Nd$_2$CuO$_4$ are presented and are compared to results obtained from NMR experiments. The results for the spin density distribution and the calculation of magnetic hyperfine couplings are discussed in Sec.~\ref{sec:spin}. Sec.~\ref{sec:summary} contains a summary and conclusions.

\section{The cluster method and computational details}

\label{sec:clustermodel}

\subsection{General Remarks}

The idea of the cluster method is to solve the many-body Schr\"odinger equation for a portion of the crystal which we call the cluster. A cluster is a careful selection of a contiguous group of ions within a solid. The specific choice of the atoms that make up a cluster is such that it allows predominantly localized properties of a target atom and its vicinity to be calculated. A property is called {\it local} if it can be determined by the electronic structure of a few neighboring atoms in the crystal. Electric field gradients (EFG) and hyperfine fields are such local properties and have been successfully calculated by the cluster method (see for example Refs.~\cite{bib:millenniumpaper,bib:ybcopaper}). A cluster consists of three regions. The target atom and at least its nearest neighboring atoms form the center of the cluster and the corresponding electrons are treated most accurately using first-principles all-electron methods. This core region is embedded in a large cloud of a few thousand point charges at the respective lattice sites imitating the Madelung potential. Point charges at the border of the core region are replaced by basis-free pseudopotentials to improve the boundary conditions for the electrons in the cluster core. These pseudopotentials make up the so-called screening region.

It would be desirable to have clusters that contain as many atoms as possible in the core region but there are two computational limitations to the cluster size: the available computer resources and the convergence of the self-consistent field procedure. However, these limits have been pushed further since our first use of the cluster technique (see Ref.~\onlinecite{bib:suter97}) which now allows the computation of larger clusters and hence improve the bulk-to-surface ratio.

\begin{table}[htb]
\begin{center}
\begin{tabular}{lccc}
\hline
  Cluster                            & $N$ & $E$   & $B$   \\ \hline
 Cu$_{5}$O$_{16}$/Cu$_{8}$Nd$_{24}$  & 21& 295 & 403 \\ 
 Cu$_{9}$O$_{24}$/Cu$_{12}$Nd$_{32}$ & 33& 483 & 663 \\
 Cu$_{13}$O$_{36}$/Cu$_{12}$Nd$_{48}$& 49& 711 & 975 \\ \hline
\end{tabular}
\end{center}
\caption{Compilation of the clusters used for \nd\ and some of their defining properties: number of atoms simulated by a full basis set ($N$), number of electrons ($E$) and number of basis functions ($B$).}
\label{tbl:compilation_nd}
\end{table}

In this work three clusters of different size (with 5, 9, and 13 copper atoms in the core region) are studied and their constitutive properties are given in Table~\ref{tbl:compilation_nd}. For clarity these clusters are labeled X/Y where X is the chemical formula of the core region and Y the formula of the screening region.

\subsection{Computational Details}

\begin{figure}[h!]
\centering

\includegraphics[width=0.48\textwidth]{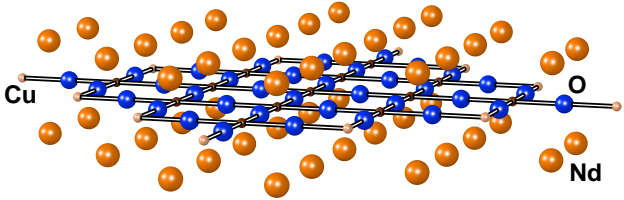}
\caption{(color online). Schematic representation of the Cu$_{13}$O$_{36}$/Nd$_{48}$Cu$_{12}$ cluster. The 13 small black spheres denote the copper atoms simulated with basis sets, the 12 small grey (brown) spheres at the border of the cluster are the coppers represented by pseudopotentials.}
\label{fig:Nd_Cu13}
\end{figure}

The atomic positions were chosen according to crystallographic structure determinations. The crystal structure of \nd\ is the so called $T'$ structure. A characteristic of this structure is the absence of apex oxygens in contrast to hole doped materials. In this case the core region consists only of planar Cu and O. The lattice parameters ($a=b=3.95$~\AA\ and $c=12.07$~\AA) and the positions of the atoms within the unit cell have been adopted from Refs.~\cite{bib:uzumaki1991,bib:xue1990}. The core region of the smallest cluster (Cu$_5$) for \nd\ incorporates five planar Cu atoms and their 16 nearest planar oxygen atoms. The screening shell includes bare pseudopotentials for 8 Cu$^{2+}$ and 24 Nd$^{3+}$ ions. The Madelung shell contains more than 6000 point charges. According to the attributed valences of copper and oxygen of 2$^+$ and 2$^-$, respectively, this cluster is given a total charge of 22$^-$, leading to a total number of 295 electrons. We will also use \cun\ and \cut\ clusters whose properties are given in Table~\ref{tbl:compilation_nd}. The largest cluster (Cu$_{13}$) is schematically depicted in Fig.~\ref{fig:Nd_Cu13}.

For the atoms in the core region the standard 6-311G basis sets~\cite{bib:szabo} were employed. The electronic structures in all the clusters described above were determined with the Gaussian03 quantum chemistry package~\cite{bib:g98} in the framework of density functional theory incorporating the exchange functional proposed by Becke~\cite{bib:becke1,bib:becke2}, together with the correlation functional of Lee, Yang, and Parr~\cite{bib:lyp} (specified by the  BLYP keyword in the Gaussian03 program).
We note that band structure calculations with the local density approximation (LDA) give no magnetic solution for Nd$_2$CuO$_4$. In cluster calculations, however, the spin state can be chosen and in all cases (irrespective of the functional, LDA or generalized gradient approximation) the state with the lowest variational energy is found for the antiferromagnetic arrangement~\cite{bib:ybcopaper,bib:martin1997}.

The results of each calculation were examined with the Mulliken population analysis which gives a description of the charge and spin densities in terms of the individual atoms and also the constituent orbitals. Other properties such as EFGs and hyperfine fields were also recorded at each copper center. It is important to realize that properties calculated with the cluster method as e.g. EFGs or hyperfine fields depend only marginally on the cluster size (see a detailed analysis in Ref.~\onlinecite{bib:renold2005}).

\subsection{Simulation of doping}
\label{sec:simdoping}

There is an important aspect to consider when studying the change of the electronic structure upon doping cuprates by means of cluster calculations. If the doping is produced by impurity atoms (as for example replacing La$^{3+}$ by Sr$^{2+}$ in \lacuo\ or Nd$^{3+}$ by Ce$^{4+}$ in \nd) one can investigate the local changes in the immediate neighborhood of the impurity atoms. In this situation the doped charge carriers are ``pinned'' around the dopant and are thus interpreted as {\it localized} holes (electrons). In contrast, we would like to investigate holes (electrons) {\it distributed} over the CuO$_2$ plane. Therefore we abstain from changing the chemical composition of our clusters but instead we simulate doping either (I) by simply removing (or adding) one electron or (II) by creating hole (electron) excess at the center of the cluster by an additional electric field. Note that in both methods, the nature of the dopant is not specified.

In this work we generalize the methods used in Refs.~\cite{bib:stollI2003,bib:stollII2003} for the hole doped material (\lacuo) to the electron-doped material (\nd). If we simulate doping by adding an electron to the cluster (method I) the spin state of the cluster changes.
The corresponding doping level is about 20~\% for a Cu$_5$ cluster, 10~\% for a Cu$_9$ cluster, and 8~\% for a Cu$_{13}$ cluster. It is therefore not possible to simulate an arbitrary doping level with this method.

For the second method we introduce additional point charges at the periphery of the cluster to create an electric field across the cluster in order to move the charge towards or away form the target atoms in the cluster center. We call this method the {\it peripheral charge method}. Note that this added system of charges has no physical meaning. It is convenient because it can be continuously altered so that the charge can be progressively directed to or extracted from the target atoms in the center of the cluster. This implies that the cluster must be sufficiently large so that ions of interest are not near the cluster edge. In contrast to the first method the cluster and the system of charges always have the same number of electrons and the same spin state, but the charge of the cluster can be progressively changed in a manner expected by doping. Of course this model has only a limited range of applicability and would ultimately break down for high doping levels since no electron is actually removed or added. The peripheral charge method serves to continuously interpolate the results already obtained with the first method (with remarkable success, see Sec.~\ref{sec:Nd}).

\section{Electronic structure}
\label{secIII} 
\subsection{Charge distribution}
\label{sec:Nd}
In cluster calculations, the charge distribution is determined by the occupied molecular orbitals (MOs). The $m^{th}$ MO is represented as a linear combination

\begin{equation}
\phi_m(\vec{r})=\sum_{K=1}^n\phi_m^K(\vec{r}-\vec{R}_K)=\sum_{K=1}^n\sum_{k=1}^{n_K}c_m^{K,k}B_{K,k}(\vec{r}-\vec{R}_K)
\label{eq:linearcomb}
\end{equation}
of $n_K$ atomic basis functions $B_{K,k}$ centered at the nuclear sites $K=1,\dots,n$, and the $c_m^{K,k}$ are the MO coefficients~\cite{bib:remark}. The charge density is then given by $\sum_m\phi_m(\vec{r})^2$ and contains both on-site and overlap terms. In the following we will use the Mulliken population analysis to describe the details of the charge distribution.

\begin{figure*}[ht]
\centering
\resizebox{0.7\textwidth}{!}{
  \includegraphics{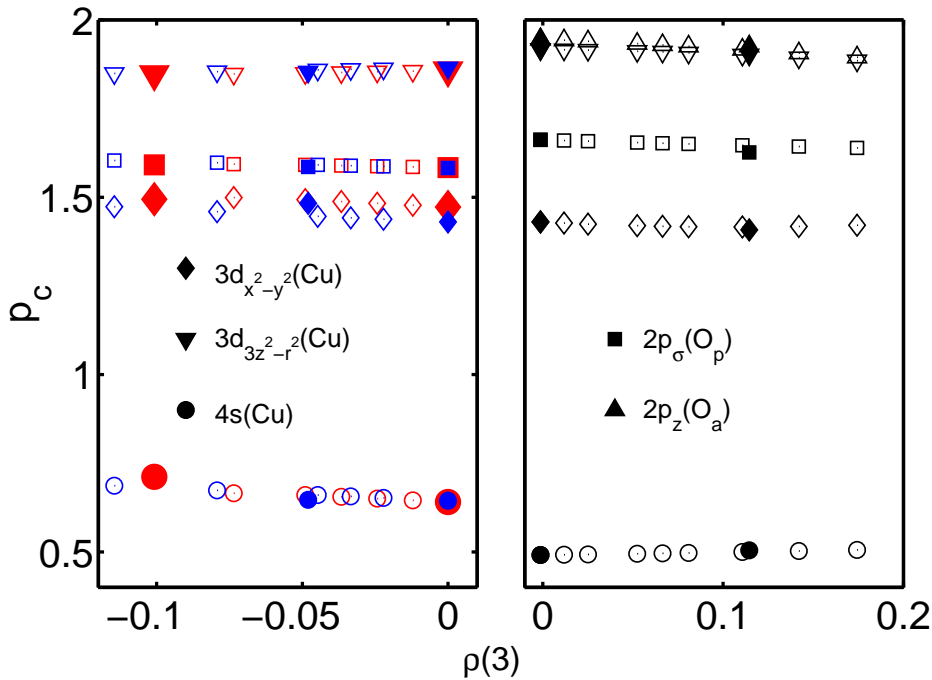}
}
\caption{(color online). Partial Mulliken populations of the atomic orbitals versus doping level $\rho(3)$ for  Nd$_{2-x}$Ce$_x$CuO$_4$ (left panel) and La$_{2-x}$Sr$_x$CuO$_4$ (right panel). The following orbitals are represented: $3d_{3z^2-r^2}$ (triangles down), 4s (circles), apical oxygen $2p_z$ (triangles up), $3d_{x^2-y^2}$ (diamonds) and the planar oxygen $2p_{\sigma}$ (squares). Open symbols refer to peripheral charge calculations and filled symbols refer to calculations where electrons are added or removed as appropriate (method I).}
\label{fig:Ndpartial}
\end{figure*}

In the left part of Fig.~\ref{fig:Ndpartial} we have represented the partial Mulliken populations, $p_c$, for the relevant orbitals: 3d$_{x^2-y^2}$, 3d$_{3z^2-r^2}$, and 4s for Cu and 2p$_\sigma$ for O$_p$. All other orbitals have populations that differ only marginally ($< 0.2 \%$) from 2. Results have been obtained from calculations with clusters of three different sizes: Cu$_{5}$O$_{16}$/Cu$_{8}$Nd$_{24}$ with spin multiplicity $M=4$, Cu$_{9}$O$_{24}$/Cu$_{12}$Nd$_{32}$ with $M=2$, and Cu$_{13}$O$_{36}$/Cu$_{12}$Nd$_{48}$ with $M=6$ for the undoped case. We find that the total energy is lowest for the multiplicity that corresponds to an antiferromagnetic spin alignment. We repeated these calculations with one additional electron and, again, with a spin multiplicity leading to the lowest ground state energy ($M=3$, 1, and 5 for Cu$_5$, Cu$_9$, and Cu$_{13}$, respectively). This doping simulation changes the total charge in the unit cell. We therefore define a quantity $\rho(3)$ as the sum of the Mulliken charge on the central copper atom and that on the adjacent oxygen atoms,
\begin{equation}
\label{rho(3)Nd}
\rho(3) = \rho(\textrm{Cu}) + 2 \rho(\textrm{O}_p) + 2,
\end{equation}
which approximately represents the fractional excess charge in the unit cell.
Ideally, in the undoped substance $\rho(3)$ should vanish. Due to open boundaries
in the cluster method this is not quite exactly the case but since all calculated 
values are sufficiently close to zero we have renormalized the values for $\rho (3)$ such that they are indeed zero in the undoped compounds.
The full large (red) symbols are the partial Mulliken populations calculated in the Cu$_5$ cluster in the undoped case ($\rho(3)=0$) and with one additional electron at $\rho(3) \approx -0.1$. The full small (blue) symbols denote $p_c$ values obtained from a cluster with 13 Cu again in the undoped case and with an additional electron at $\rho(3) \approx -0.05$. The results of the calculation for the cluster with 9 atoms are not shown although in the undoped case they are almost identical to those of Cu$_{13}$. However, when adding an electron, the results exhibit a peculiar superstructure for which presently we do not have a unique explanation. The open symbols, finally, represent partial populations that have been determined with the peripheral charge method described in Sec.~\ref{sec:simdoping} that allows to simulate the doping with fractional charges. These values have been obtained with the Cu$_5$ (Cu$_{13}$) cluster and result from calculations that all used the same spin multiplicity $M=4$ ($M=6$) as in the undoped case.
(Note that the partial populations of the 3d$_{3z^2-r^2}$, 4s, and 2p$_{\sigma}$ orbitals in the Cu$_5$ and Cu$_{13}$ clusters for $\rho(3)=0$ are so close  that the corresponding markers nearly completely overlap.)
It is seen that in the undoped case the partial populations only marginally depend on the cluster size. As a function of doping the $p_c$ values calculated for the two differently sized clusters (filled symbols) change monotonically and are reasonably well interpolated by those obtained with the peripheral charge method (open symbols).

For a general discussion it is instructive to compare these results with those obtained previously by Stoll {\it et al.}~\cite{bib:stollI2003} for hole doped La$_{2-x}$Sr$_x$CuO$_4$ which are shown in the right part of Fig.~\ref{fig:Ndpartial}. Note that in this case $\rho(3)$ is defined by
\begin{equation}
\label{rho3}
 \rho(3)=\rho({\rm{Cu}}) + 2 \rho({\rm{O}}_p)+2\rho({\rm{O}}_a)+
6.
\end{equation}

Again, the full symbols denote partial Mulliken populations of the relevant AO calculated for the undoped material $(\rho(3) = 0)$ for the cluster with 5 Cu atoms with $M=4$ and with one electron less and $M=3$ ($\rho(3) \approx 0.12$). The open symbols are obtained from calculations with the peripheral charge method.\\

By comparing the left and right sides of Fig.~\ref{fig:Ndpartial} it is first observed that there are no drastic differences between the electron and hole doped materials although the latter have apical oxygens whose 2p$_z$ orbitals are not completely filled. The populations of the Cu 3d$_{x^2-y^2}$ orbitals (diamonds) and the O 2p$_\sigma$ orbitals (squares) which carry most of the spin density slightly decrease from left to right. There is of course a discontinuity at $\rho(3) = 0$ which is due to the different structures (the lattice constant $a$ for Nd$_2$CuO$_4$ is 4.8~\% larger than for La$_2$CuO$_4$).

\begin{figure}[htb]
\centering
\resizebox{0.48\textwidth}{!}{
  \includegraphics{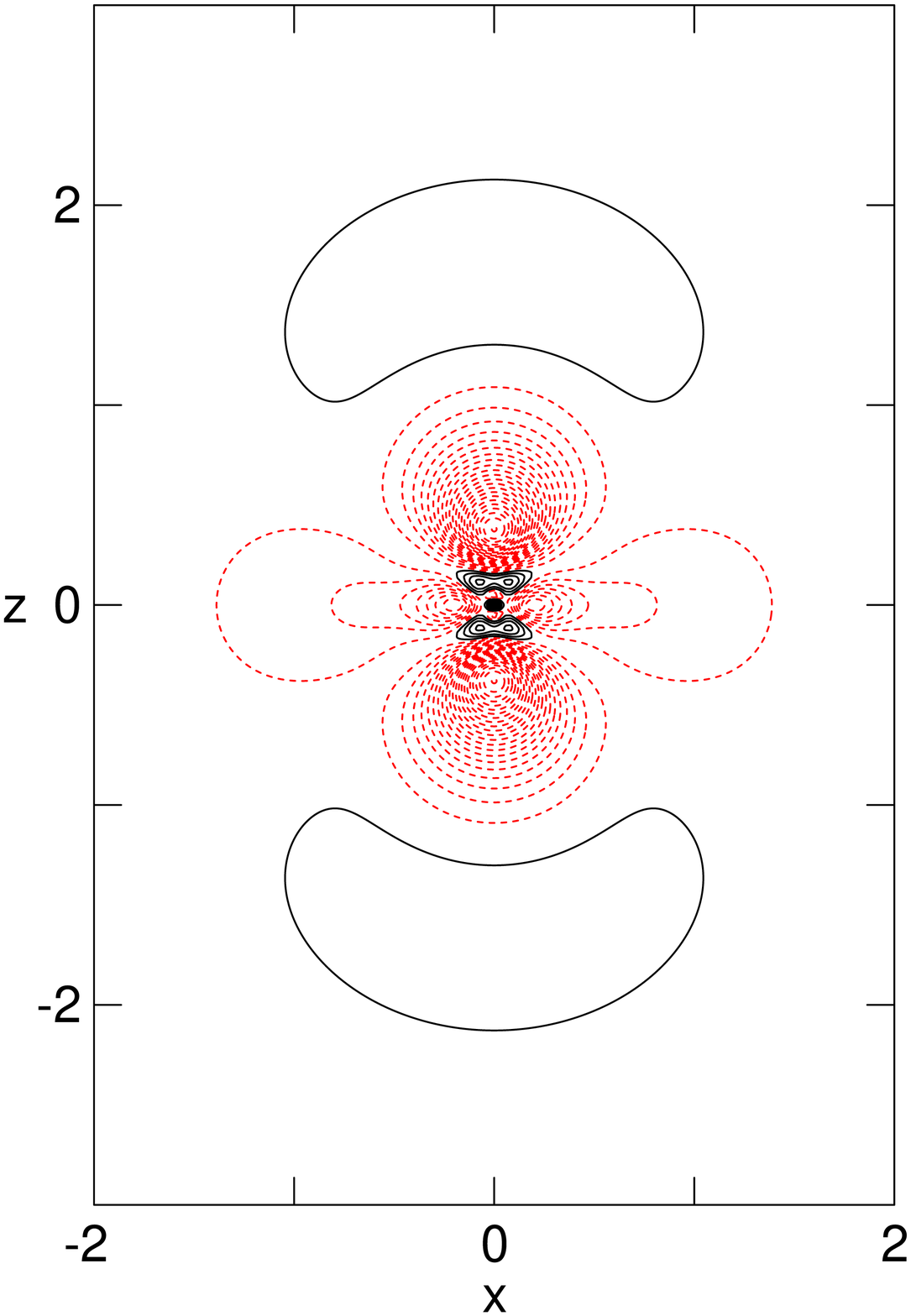}
}
\caption{(color online). Electron density difference contour plots in the x-z plane (perpendicular to the CuO$_2$ plane) between the Cu$_{13}$O$_{36}$/Cu$_{12}$Nd$_{48}$ and the Cu$_{13}$O$_{62}$/Cu$_{12}$La$_{74}$ clusters on a area of 4 a$_B$ $\times$ 6 a$_B$. The wave functions of the cluster electrons have been projected onto the 4s and the 3d$_{3z^2-r^2}$ states. The solid density lines (black) occur when the local electron density of the cluster containing Nd is larger than the electron density of the cluster containing La. Dashed lines (red), on the other hand, denote that the local electron density is larger in the La containing cluster. Two neighboring lines (solid and dashed) are separated by an electron density difference of 0.01~$e$/a$_B^3$.}
\label{fig:densdiffNdLa}
\end{figure}

It is remarkable that the Cu 3d$_{3z^2-r^2}$ AO (triangles down) in the electron-doped material is less occupied than in La$_2$CuO$_4$. This is compensated by a significantly larger population of the Cu 4s AO (circles). In Fig.~\ref{fig:densdiffNdLa} the density difference in the 4s and \dz\ AOs between the  Cu$_{13}$O$_{36}$/Cu$_{12}$Nd$_{48}$ and Cu$_{13}$O$_{62}$/Cu$_{12}$La$_{74}$ is shown which demonstrates the increased population of the 4s at the expense of the \dz\ AO.\\

\subsection{Sensitivity of EFG values on charge distribution}
\label{sec:sensitivity}

A detailed description of the evaluation of EFGs in the framework of cluster calculations is found in Ref.~\onlinecite{bib:stoll}. Here we just point out the relevant aspects that are connected to the rearrangement of charge distributions due to electron doping.

We assume in the following that the target nucleus $K_T$ is at $\vec{R}_{K_T}=0$. The contribution of the MO $\phi_m$ to the EFG at $K_T$ is given by the matrix element
\begin{equation}
V_m^{ij}=\dotp{\phi_m(\vec{r})}{\frac{3x^ix^j-r^2\delta^{ij}}{r^5}}{\phi_m(\vec{r})}
\nonumber
\end{equation}
\begin{equation}
=\sum_{K=1}^{n}\sum_{L=1}^{n}\sum_{k=1}^{n_K}\sum_{l=1}^{n_L}c_m^{K,k}c_m^{L,l} \times
\label{eq:matrixel}
\nonumber
\end{equation}
\begin{equation}
\times \dotp{B_{K,k}(\vec{r}-\vec{R}_K)}{\frac{3x^ix^j-r^2\delta^{ij}}{r^5}}{B_{L,l}(\vec{r}-\vec{R}_L)}.
\end{equation}
This matrix element contains contributions from basis functions centered at two nuclear sites $K$ and $L$. Thus we can identify three types of contributions: $(i)$ on-site terms from basis functions centered at the target nucleus ($K=L=K_T$, contribution I), $(ii)$ mixed on-site and off-site contributions (II), and $(iii)$ purely off-site terms with $K\neq K_T$ and $L \neq K_T$ (III).

In addition, there is a contribution coming from all nuclear point charges $Z_K$ which we denote by W$_{ij}$. The charges of the bare nuclei at sites $K \ne K_T$ are screened by the matrix elements with $K = L$. Therefore the combined contributions from III and the nuclei (W$_{ij}$) are small. The partitioning of the contributions to the EFG tensor $V_{ij}$ thus reads
\begin{equation}
V_{ij}={^IV_{ij}}+{^{II}V_{ij}}+{^{III}V_{ij}}+W_{ij}.
\end{equation} 
 More details about these regional partitions can be found in Ref.~\onlinecite{bib:stoll}.

\begin{table}[htb]
\begin{center}

\begin{tabular}{llrr}
& & undoped & electron-doped \\
& & ($M=6$) & ($M=5$)        \\ \hline
I  & p                  & $-1.431$ & $-1.433$ \\
   & d$_{x^{2}-y^{2}}$  & $-6.378$ & $-6.613$ \\
   & d$_{xy}$           & $-9.128$ & $-9.088$ \\
   & d$_{3z^{2}-r^{2}}$ &   8.309  &   8.222  \\
   & d$_{xz}$           &   4.528  &   4.512  \\
   & d$_{yz}$           &   4.528  &   4.512  \\ \hline
II & s                  &   0.374  &   0.396  \\
   & p                  &   0.013  &   0.013  \\
   & d$_{x^{2}-y^{2}}$  & $-0.021$ & $-0.020$ \\
   & d$_{xy}$           &   0.001  &   0.001  \\
   & d$_{3z^{2}-r^{2}}$ & $-0.207$ & $-0.217$ \\
   & d$_{xz}$           &   0.000  &   0.000  \\
   & d$_{yz}$           &   0.000  &   0.000  \\ \hline
\multicolumn{1}{l}{III + Nuclei}     &  & $-0.015$ & $-0.017$ \\ \hline
\multicolumn{1}{l}{Point charges}    &  & $-0.033$ & $-0.033$ \\ \hline
\multicolumn{1}{l}{Total}            &  &   0.538  &   0.235  \\
\end{tabular}
\caption{Contributions I, II, III to the EFG component $V_{zz}$ for the central copper (in atomic units), calculated for the Cu$_{13}$ cluster representative of \nd.
  }
\label{tab:contr2}
\end{center}
\end{table}

We collect in Table~\ref{tab:contr2} the various contributions to $V_{zz}$ at the copper obtained for the Cu$_{13}$O$_{36}$/Cu$_{12}$Nd$_{48}$ cluster for both the undoped and the electron-doped calculations. In the undoped case, the total value $V_{zz} = 0.538 $ is obtained from contributions I (0.428), II (0.160), III + nuclei ($-0.015$) and a small correction due to the surrounding point charges ($-0.033$). In the electron-doped case the total EFG of $V_{zz} = 0.235$ is substantially smaller which is mainly due to a diminished contribution from region I.
 Focusing on region I we note that the added contributions from the three AO 3d$_{xy}$, 3d$_{zx}$, and 3d$_{yz}$ are $-0.072$ (undoped) and $-0.064$ (doped).
 
A careful analysis of Table~\ref{tab:contr2} shows that the most significant 
difference is due to the d$_{x^{2}-y^{2}}$ and the d$_{3z^{2}-r^{2}}$ parts
of contribution I. Only very small corrections are obtained by the other 
contributions. The influence of part III of the electrons on the other atoms
is nearly fully compensated by the nuclear charges of these atoms and 
contributes only by a very small amount. 

To understand the changes of the EFG on the copper sites upon doping, it is therefore sufficient to concentrate on the MOs which contain partially occupied  3d$_{x^2-y^2}$ and 3d$_{3z^2-r^2}$ AOs at the target nucleus.
The EFG is then given by
\begin{equation}
V_{zz}=R+\frac{4}{7} ( N_{3z^2-r^2}\expect{r^{-3}}_{3z^2-r^2} - N_{x^2-y^2}\expect{r^{-3}}_{x^2-y^2}),
\label{eq:EFG}
\end{equation}
where $R$ is the contribution from all other orbitals and $N$ are the partial occupation numbers of the corresponding orbitals which are gathered in Table~\ref{tbl:occupancies}.

The fact that the partial occupation numbers $N$ of the AOs are very similar to the partial Mulliken populations $p_c$ which are also included in Table~\ref{tbl:occupancies} gives evidence that despite their simplistic concept the Mulliken charges can be significant quantities in the analysis of charge distributions in cuprates. In particular, the differences $N_{3z^2-r^2}-N_{x^2-y^2}$ and $p_c(3d_{3z^2-r^2})-p_c(3d_{x^2-y^2})$ are very similar. In Sec.~\ref{sec:EFG} we will therefore use a simplified version of Eq.~(\ref{eq:EFG}) for a more general analysis of the Cu EFGs in cuprates.

\subsection{Comparison with experiments}

The nuclear quadrupole frequency, $\nu_Q$, is related to the main component of the EFG, $V_{zz}$, by

\begin{equation}
\nu_Q=\frac{3e}{2h}\frac{1}{I(2I-1)}QV_{zz}\sqrt{1+\eta^2/3}.
\end{equation}
Here, $Q$ is the quadrupole moment of the nucleus under consideration and $\eta$ is the anisotropy parameter.
For $I=3/2$ and for axial symmetry ($\eta=0$) one has
\begin{equation}
\nu_Q=\frac{e}{2h}QV_{zz}.
\end{equation}
If we adopt the value of the quadrupole moment $^{63}Q=-0.211$~b derived by Sternheimer~\cite{bib:sternheimer} we obtain from our calculations a quadrupolar frequency of $^{63}\nu_Q=13.3$~MHz in the undoped Nd$_2$CuO$_4$ which is in very good agreement with the value of $\nu_Q=14$~MHz reported by Abe {\it et al.}\cite{bib:abe1989}. In the electron-doped materials we get $^{63}\nu_Q=5.8$~MHz in the Cu$_{13}$ cluster (corresponding to a doping level of about 8~\%) and $^{63}\nu_Q=3.6$~MHz in the Cu$_5$ cluster (corresponding to 20~\% doping). Experimental values for quadrupole frequencies in doped materials are generally much smaller than in the undoped case and close to zero ($\nu_Q<2$ MHz in Ref.~\onlinecite{bib:kambe1991} for a doping level of 15~\%). This reduction compared to the undoped substance is at least qualitatively reproduced by our calculations.

\begin{table}[htb]
\begin{center}
\begin{tabular}{lcccc}
\hline
& \multicolumn{2}{c}{undoped} & \multicolumn{2}{c}{electron-doped} \\
& $3d_{x^2-y^2}$ & $3d_{3z^2-r^2}$ & $ 3d_{x^2-y^2}$ & $3d_{3z^2-r^2}$ \\ \hline
$N$                & 1.379 & 1.820 & 1.435 & 1.806 \\
$p_c$              & 1.430 & 1.865 & 1.483 & 1.856 \\
$\expect{r^{-3}}$  & 8.094 & 7.991 & 8.064 & 7.968 \\

\hline\end{tabular}
\caption{Partial occupation numbers $N$, partial Mulliken populations $p_c$, and expectation values $\expect{r^{-3}}$ for the relevant copper AOs in undoped ($M=6$) and doped ($M=5$) Cu$_{13}$ clusters representative of Nd$_2$CuO$_4$.}
\label{tbl:occupancies}
\end{center}
\label{tbl:MullikenchargediffYBCO}
\end{table}

\subsection{General aspects of Cu EFG}

\label{sec:EFG}

\begin{figure}[htb]
\centering
\resizebox{0.48\textwidth}{!}{
  \includegraphics{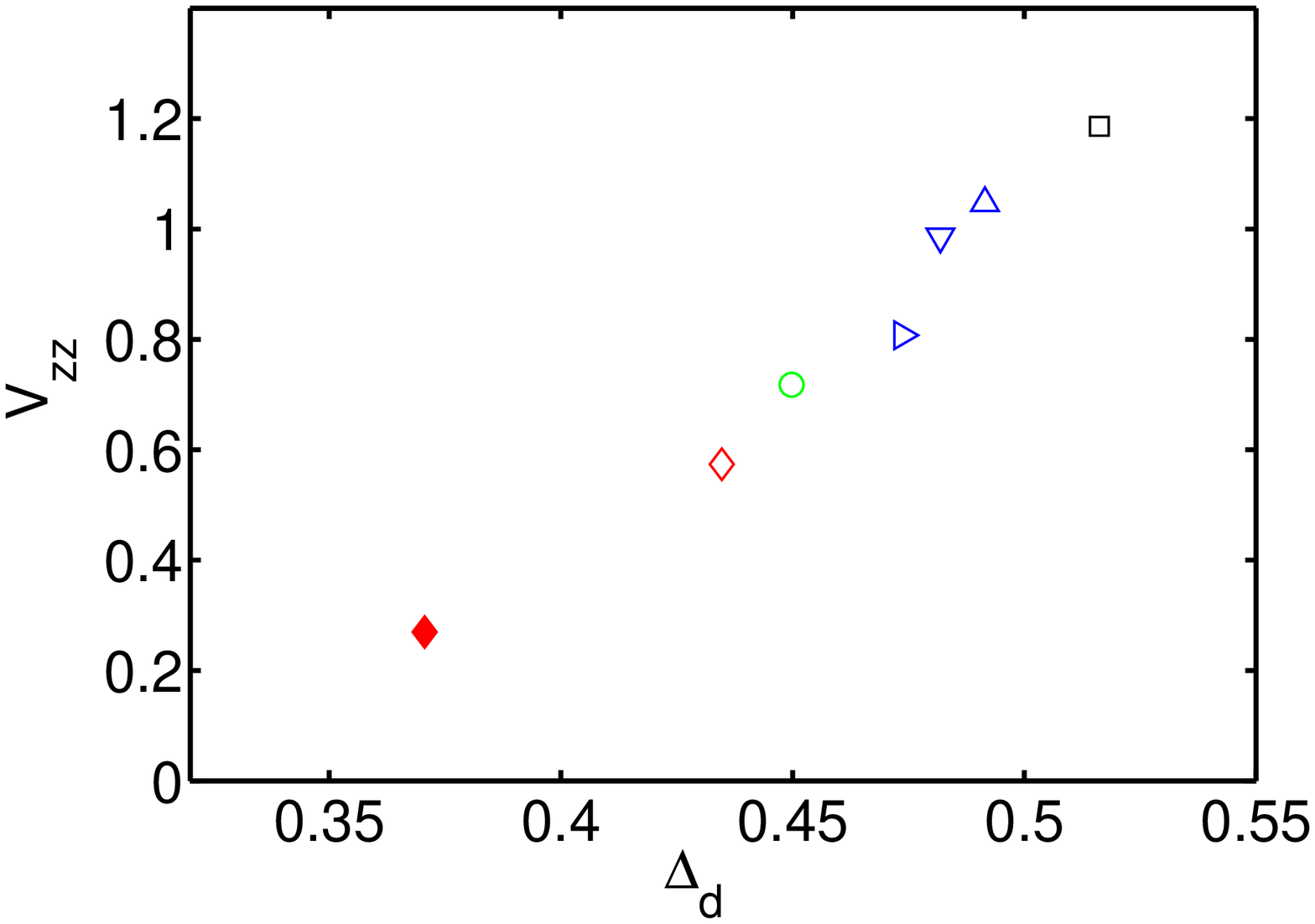}
}
\caption{(color online). EFG versus $\Delta_d$=$p_c$(3d$_{3z^2-r^2}$)$-p_c$(3d$_{x^2-y^2}$) for \lacuo\ (black square), \ybco{7} (blue triangle up), \ybc\ (blue triangle down), \ybco{6} (blue triangle right), \srcl\ (green circle), \nd\ (red diamond), Nd$_{1.92}$Ce$_{0.08}$CuO$_4$ (full red diamond).}
\label{fig:EFG_pc}
\end{figure}

It is often assumed (by making use of the simple ionic model) that the EFG can be used to give an estimate of the partial Mulliken population, $p_c$(3d$_{x^2-y^2}$), of the Cu 3d$_{x^2-y^2}$ orbital. This is of course only true if one assumes that all the other d orbitals are fully occupied. If the population, $p_c$($3d_{3z^2-r^2}$), of the Cu 3d$_{3z^2-r^2}$ orbital is also less than 2, Eq.(\ref{eq:EFG}) must be used for the description of the EFG. If we replace the occupation numbers in Eq.(\ref{eq:EFG}) by the partial Mulliken populations and if we assume that $\expect{r^{-3}}$ is similar for all 3d orbitals we arrive at the following simplified formula:

\begin{equation}
V_{zz} \approx \frac{4}{7}\langle r^{-3}\rangle \Delta_d + R.
\label{eq:vzz}
\end{equation}
Here $\Delta_d=p_c(3d_{3z^2-r^2})-p_c(3d_{x^2-y^2})$. In Table~\ref{tbl:laynd} we have collected the relevant partial Mulliken populations $p_c$($3d_{x^2-y^2}$) and $p_c$($3d_{3z^2-r^2}$) together with $\Delta_d$ and  $V_{zz}$ for several cuprates (including \ybco{6}\ (Ref.\onlinecite{bib:renold2005}), \ybco{7}\ (Ref.~\onlinecite{bib:ybcopaper}), \ybc\ (Ref.~\onlinecite{bib:renold2005}), and \srcl\ (Ref.~\onlinecite{bib:bersier2002})). For further reference Table~\ref{tbl:laynd} also contains the 4s populations. In Fig.~\ref{fig:EFG_pc} we have plotted $V_{zz}$ versus $\Delta_d$ for various cuprates. The linear correlation is evident and we thus conclude that the changes in $V_{zz}$ depend almost entirely on the population of the 3d$_{x^2-y^2}$ and the 3d$_{3z^2-r^2}$ orbitals. This modification of the ionic model tells us that a small value of $V_{zz}$ can also be obtained by an appropriate population difference of the \dz\ and \dx\ AOs.

\begin{table}[htb]
\begin{center}
\begin{tabular}{l|ccc|cc}
\hline
Substance & \multicolumn{3}{c|}{Mulliken populations}&&\\
\cline{2-4}
                              & \dx    & \dz    &  4s    & $\Delta_d$ & V$_{zz}$   \\ \hline
\lacuo                        & 1.406 & 1.922 & 0.502 & 0.516 & 1.186 \\
\nd                           & 1.430 & 1.865 & 0.647 & 0.435 & 0.574 \\
 Nd$_{1.92}$Ce$_{0.08}$CuO$_4$& 1.483 & 1.854 & 0.644 & 0.371 & 0.270 \\
\ybco{7}                      & 1.423 & 1.914 & 0.590 & 0.492 & 1.047 \\
\ybc                          & 1.427 & 1.909 & 0.592 & 0.482 & 0.986 \\
\ybco{6}                      & 1.428 & 1.902 & 0.618 & 0.474 & 0.807 \\
\srcl                         & 1.450 & 1.899 & 0.525 & 0.450 & 0.718 \\ 
\hline

\end{tabular}
\end{center}
\caption{Partial Mulliken populations of the \dx\ and \dz\ orbitals, their difference $\Delta_d$, and the calculated EFG for \lacuo, undoped and electron-doped \nd, three substances of the YBaCuO family, and \srcl\ for Cu$_{13}$ clusters.}

\label{tbl:laynd}
\end{table}

The relationship between the calculated and the {\it measured} EFGs will be discussed in depth in a subsequent paper\cite{bib:renold2005}.

\subsection{Orbital Shifts}
\label{sec:orbital}
For the cluster Cu$_{5}$O$_{16}$/Cu$_8$Nd$_{24}$ we also have determined the chemical shielding tensor for the central copper nucleus in the same way as has been reported previously for La$_2$CuO$_4$ in Ref.~\onlinecite{bib:renold03}. The isotropic diamagnetic shielding is $^{63}\sigma_{dia}=2400$ ppm. This value is within a few ppm the same for all Cu compounds. The paramagnetic shieldings are evaluated as $^{63}\sigma^{\parallel}_{para}=-(8000\pm 800)$~ppm and  $^{63}\sigma^{\perp}_{para}=-(2600\pm 300)$~ppm for the field directions parallel and perpendicular to the c-axis, respectively. The error bars are estimated from the dependencies of the size of the cluster.
To compare these theoretical values with measurements we need to reassess the analysis of orbital shift measurements. As has been discussed in detail in Ref.~\onlinecite{bib:renold03}, the NMR shifts for a nuclear species $k$ are commonly determined as differences of frequencies measured in the same magnetic field in the target (t) substance and some reference (r) substance. Their connection with the chemical shieldings is given by
\begin{equation}
{^kK_L^{ii}(t-r)}={^k\sigma_{dia}^{ii}(r)}+{^k\sigma_{para}^{ii}(r)}-{^k\sigma_{dia}^{ii}(t)}-{^k\sigma_{para}^{ii}(t)}.
\end{equation} 
For the copper nucleus, the reference substance is usually CuCl and in the above equation the diamagnetic terms cancel out:
\begin{equation}
{^{63}K_L^{ii}(t-{\rm CuCl})}={^{63}\sigma_{para}^{ii}({\rm CuCl})}-{^{63}\sigma_{para}^{ii}(t)}.
\end{equation}
The paramagnetic shielding for CuCl, however, is substantial and has experimentally been determined by Lutz {\it et al.}~\cite{bib:LutzI,bib:LutzII} to be $^{63}\sigma_{para}^{iso}({\rm CuCl})=-1500$ ppm with an error of about 10 ppm. As it was done in Ref.~\onlinecite{bib:renold03} we introduce the notation of paramagnetic field modification $^k\overline{K}^{ii}_L={-^k\sigma_{para}^{ii}}$. For copper in Nd$_2$CuO$_4$ we get
\begin{equation}
^{63}\overline{K}^{\parallel}_L=(0.80\pm0.08)\%\quad\textrm{and}\quad ^{63}\overline{K}^{\perp}_L=(0.26\pm0.03)\%
\end{equation}
We are not aware of data for orbital shifts in doped Nd$_2$CuO$_4$. Detailed measurements, however, have been reported by Zheng {\it et al.}~\cite{bib:zheng2003} on Pr$_{0.91}$LaCe$_{0.09}$CuO$_{4-y}$. The deduced orbital shifts are $^{63}K^{\parallel}_L=0.92\%$ and  $^{63}K^{\perp}_L=0.155\%$, measured relatively to the reference substance CuCl. After correction for $^{63}\sigma^{iso}_{para}$(CuCl), as discussed above, one obtains the experimental paramagnetic field modifications $^{63}\overline{K}^{\parallel}_L=1.07\%$ and  $^{63}\overline{K}^{\perp}_L=0.305\%$, respectively. The agreement of the theoretical value for the perpendicular component with the experiment is reasonably good. The predicted value for $^{63}\overline{K}^{\parallel}_L$, however, is much smaller than that deduced from the data. The same discrepancy is also observed for hole doped cuprates. We note, however, that the experimental value for $^{63}\overline{K}^{\parallel}_L$ is based on the observation of a temperature independent total magnetic shift and the assumption that the on-site and transferred hyperfine fields cancel out in Pr$_{0.91}$LaCe$_{0.09}$CuO$_{4-y}$, as they do in the YBaCuO and LaSCO families. As will be discussed in Sec.~\ref{sec:spin} our calculated values of the hyperfine couplings give no support for these incidental cancellations in substances with appreciable differences in the lattice parameters.

\section{Spin density distribution and hyperfine couplings}
\label{sec:spin}

\begin{figure}[ht]
\centering
\resizebox{0.48\textwidth}{!}{
  \includegraphics{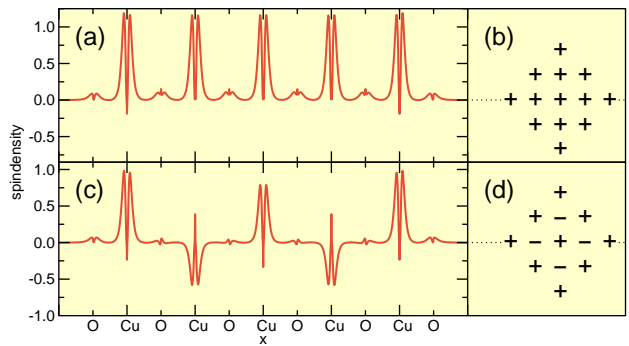}
}

\caption{(color online). (a) Spin density along the 5 Cu and 6 O in the Cu$_{13}$O$_{36}$/Cu$_{12}$Nd$_{48}$ cluster of spin multiplicity $M=14$; (c) spin density with  multiplicity $M=6$. (b) and (d) signs of the Mulliken spin densities.}
\label{fig:spind}
 \end{figure}

The spin-polarized cluster calculations allow us to investigate the spin density distribution in detail. In Fig.~\ref{fig:spind}a we depict the spin density distribution along the five Cu and six O obtained for the Cu$_{13}$O$_{36}$/Cu$_{12}$Nd$_{48}$ cluster with spin multiplicity $M=14$, which corresponds to a ferromagnetic alignment of the copper moments represented in Fig.~\ref{fig:spind}b. Most of the spin density is provided by the singly occupied molecular orbital which is also highest in energy. It is a linear combination of 3d$_{x^2-y^2}$ atomic orbitals (AOs) on the coppers ($\approx 80 \%$) and 2p$_\sigma$ AOs on the oxygens ($\approx 20 \%$). The square of the 3d$_{x^2-y^2}$ AO has maxima at distances of 0.35 {\AA} from the nucleus but vanishes at the nucleus. In general, the spin density at a nuclear site is due to s-type AOs giving rise to the Fermi contact term. In particular, the spin densities at the Cu nuclei (cusps in Fig.~\ref{fig:spind}a) can be correlated with the Mulliken spin densities (whose signs are given in Fig.~\ref{fig:spind}b) at the same and the nearest neighbor Cu nuclei. This observation can be used to split the Fermi contact term into on-site and transferred terms as has been discussed in detail in Refs.~\cite{bib:millenniumpaper,bib:ybcopaper}. We just note here that the on-site term is negative if the Mulliken spin density at the Cu atom under consideration is positive and vice versa and that the transferred term is positive (negative) if the Mulliken spin densities at the neighboring Cu atoms is positive (negative). Near the oxygens, the spin density is provided by the 2p$_\sigma$ AO and the small cusp at the position of the four innermost oxygen nuclei is due to the transferred hyperfine fields from the electronic moments of the two adjacent copper atoms.
In Fig.~\ref{fig:spind}c we show the spin density distribution obtained with spin multiplicity $M=6$. For this multiplicity, the total energy is lower than for all other multiplicities and corresponds to an antiferromagnetic alignment of the copper moments represented in Fig.~\ref{fig:spind}d.

From the differences in the total energy of the various spin configurations it is possible to estimate the antiferromagnetic exchange interaction $J$ as will be described elsewhere~\cite{bib:renold2005}. Preliminary results indicate $J \approx 160$~meV for Nd$_2$CuO$_4$ which is about the same as in La$_2$CuO$_4$ and slightly larger than in YBa$_2$Cu$_3$O$_7$ ($J \approx 130$~meV).

\begin{figure}[ht]
\centering
\resizebox{0.48\textwidth}{!}{
  \includegraphics{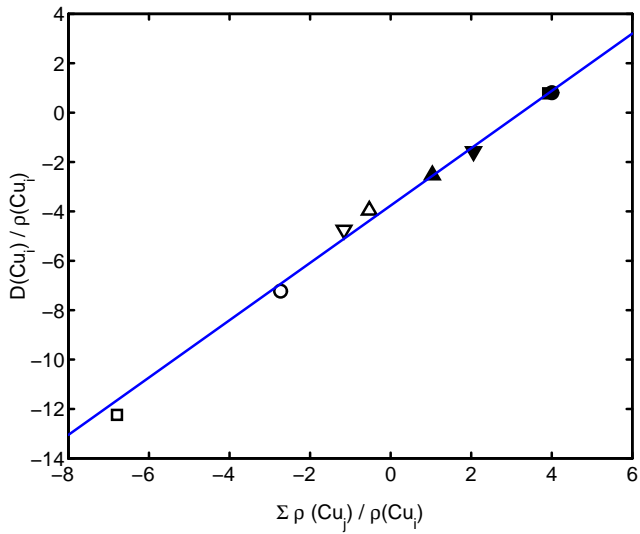}
}

\caption{
$D(\textrm{Cu}_i) / \rho(\textrm{Cu}_i)$ plotted against $\sum_{j \in NN}$ $\rho(\textrm{Cu}_j)$ / $\rho(\textrm{Cu}_i)$. Values for the cluster with multiplicity $M=14$ ($M=6$) are plotted with filled (open) symbols. The different markers correspond to symmetrically inequivalent copper sites in the cluster which differ e.g. by their number of nearest neighbors.}
\label{fig:hyperfine}
 \end{figure}

The hyperfine coupling tensor at the copper is determined by an on-site term $a^{ii}_{tot}$ and a transferred term $b^{ii}_{tot}$ from the nearest neighbors. The value of $a_{tot}^{ii}$ is made up of three contributions, an isotropic on-site interaction, $a_{iso}$, a dipolar interaction, $a_{dip}^{ii}$, and a contribution from spin-orbit coupling, $a_{so}^{ii}$. The transferred term, $b_{tot}^{ii}$, is dominated by the isotropic interaction, $b_{iso}$, since the dipolar term, $b_{dip}^{ii}$, is small and the influence of spin-orbit coupling is entirely neglected. In Fig.~\ref{fig:hyperfine} we show the dependence of the isotropic hyperfine density, $D_{iso}$, at the Cu nucleus on the on-site atomic spin density, $\rho_i$, and those of the nearest neighbors, $\rho_j$. The straight line is a fit to the equation
\begin{equation}
D_{iso}=a_{iso}\cdot\rho_i+\sum_j b_{iso}\cdot\rho_j
\end{equation}
which allows the determination of $a_{iso}$ and $b_{iso}$. Analogously, the dipolar contributions can be determined. The resulting hyperfine couplings are collected in Table~\ref{tbl:hyperfine} and are compared to the values obtained for La$_2$CuO$_4$. The total values $a^{\parallel}_{tot}$ and $a^{\perp}_{tot}$ include spin-orbit terms $a^{\parallel}_{so} = 2.222$~a$_B^{-3}$ and $a^{\perp}_{so} = 0.394$~a$_B^{-3}$ which have been estimated according to the procedure given in Ref.~\onlinecite{bib:millenniumpaper}. We note that the values for the two materials are very similar. The only significant change is in $a^{\perp}_{tot}$, which shows a different sign in the two substances.

From experiments there is much less information available for the hyperfine couplings in Nd$_2$CuO$_4$ than in the hole-doped materials. Zheng {\it et al.}~\cite{bib:zheng2003} reported spin shift measurements in Pr$_{2-x}$Ce$_x$CuO$_4$ and found, with the field parallel to the c axis, the same temperature independent behavior as has been observed in doped La$_2$CuO$_4$ and in the YBaCuO family. The spin shifts measured with field in the plane, however, change with decreasing temperature. This behavior is commonly explained by a coincidental cancellation of the hyperfine coupling contributions in c direction. The theoretical values for $a^{\parallel}_{tot}+4b^{\parallel}_{tot}$ are indeed close to zero. It is intriguing, however, that the same precise cancellation is required in most hole doped materials and in Pr$_{2-x}$Ce$_x$CuO$_4$ is intriguing to explain the temperature independent spin shifts.
More information on hyperfine couplings is available for the electron-doped infinite layer compound SrCuO$_2$. There, $a^{\perp}_{tot}+4b^{\perp}_{tot}$ should vanish to explain the data. Our calculated values give no support for these observations. 

\begin{table}[htb]
\begin{center}
\begin{tabular}{lrr}
\hline

 & La$_2$CuO$_4$& Nd$_2$CuO$_4$ \\ \hline
 $a_{iso}$              & $-$1.94 & $-$2.42 \\
 $a_{dip}^{\parallel}$  & $-$3.55 & $-$3.38 \\
 $b_{iso}$              &    0.77 &    0.75 \\
 $b^{\parallel}_{dip}$  &    0.08 &    0.03 \\
 $a^{\parallel}_{tot}$  & $-$3.30 & $-$3.58 \\
 $a^{\perp}_{tot}$      &    0.22 & $-$0.34 \\
\hline\end{tabular}
\caption{Comparative table of the hyperfine constants for La$_2$CuO$_4$ and Nd$_2$CuO$_4$ (in units of a$_B^{-3}$).}
\end{center}
\label{tbl:hyperfine}
\end{table}

\section{Summary and conclusions}
\label{sec:summary}         
Spin-polarized ab-initio calculations have been performed to determine the local electronic structure of Nd$_2$CuO$_4$ using clusters comprising 5, 9, and 13 copper atoms in the CuO$_2$ plane. Electron doping has been simulated by two different approaches which both yield the same results. The local charge distribution has been discussed in detail in terms of the partial Mulliken populations of the individual atomic orbitals. The changes in the partial populations of the relevant AOs (\dx, \dz, and $2p_{\sigma}$) that occur upon doping are relatively small but significant. Both $p_c$(\dx) and $p_c$($2p_{\sigma}$) slightly increase and $p_c$(\dz) is significantly less than 2.
Electric field gradients depend sensitively on the non-spherical local charge distribution. The individual contributions to the EFG have been discussed in detail and the resulting value for the nuclear quadrupole frequency is in good agreement with the experiments. The comparatively small value and the changes upon doping have been explained by the differences in the occupation numbers $N_{x^2-y^2}$ and $N_{3z^2-r^2}$, which are close to the partial Mulliken populations $p_c$.
The local electronic structure of Nd$_2$CuO$_4$ has been compared to that of La$_2$CuO$_4$. Hole doping in the latter material mainly decreases the population $p_c$(\dx) and $p_c$($2p_{\sigma}$) but there is also a slight decrease of $p_c$(\dz) and $p_c(2p_{z})$ of the apical oxygen. The different values in the EFG at the copper sites observed in the various cuprates can be understood by the difference in the partial Mulliken population $p_c$(\dz)$-p_c$(\dx).
The calculated spin density distribution in Nd$_2$CuO$_4$ is not much different from that determined for La$_2$CuO$_4$ and YBa$_2$CuO$_7$. The analysis of the magnetic hyperfine couplings, which is based on the dependence of the contact and dipolar spin densities also yields values which are of similar magnitudes to those in hole doped materials.  
\begin{acknowledgments}
Special thanks go to M. Mali and J. Roos for numerous stimulating discussions. Parts of this work were done in collaboration with T. A. Claxton whose help and critical discussions are acknowledged. We also thank A.-C. Uldry for the careful proof-reading of the manuscript. This work was partially supported by the Swiss National Science Foundation.
\end{acknowledgments}



\begin{thebibliography}{99}
\bibitem{bib:remark}
For the present discussion we do not need to distinguish the spin orbitals.
\end{thebibliography}
\end{document}